\documentclass[12pt,twoside,a4paper]{article}

\usepackage{amsmath,amsfonts,bm, amssymb,mathtools}

\usepackage[numbers]{natbib}

\usepackage{algorithm}
\usepackage{algpseudocode}

\usepackage{graphicx}
\usepackage{subcaption}
\usepackage{url}

\usepackage[toc]{appendix}

\begin{document}

\title{\bf A Pseudo-Marginal Metropolis-Hastings Algorithm for Estimating Generalized Linear Models in the Presence of Missing Data }
\author{
	Taylor R. Brown \hspace{.2cm}\\
    Department of Statistics \\
    University of Virginia
	\and
	Timothy L. McMurry \hspace{.2cm}\\
    School of Medicine \\
    University of Virginia
    \and 
    Alexander Langevin \hspace{.2cm}\\
    Department of Systems and Information Engineering \\
    University of Virginia
}
\maketitle

\bigskip
\begin{abstract}
The missing data issue often complicates the task of estimating generalized linear models (GLMs). We describe why the pseudo-marginal Metropolis-Hastings algorithm, used in this setting, is an effective strategy for parameter estimation. This approach requires fewer assumptions, it provides joint inferences on the parameters in the likelihood, the covariate model, and the parameters of the missingness-mechanism, and there is no logical inconsistency of assuming that there are multiple posterior distributions. Moreover, this approach is asymptotically exact, just like most other Markov chain Monte Carlo techniques. We discuss computing strategies, conduct a simulation study demonstrating how standard errors change as a function of percent missingness, and we use our approach on a ``real-world" data set to describe how a collection of variables influences the car crash outcomes.
\end{abstract}


\newpage

\section{Introduction}

In this article, we investigate a pseudo-marginal Metropolis-Hastings algorithm for handling missing data in the context of logistic regression. Missing data is a common problem in many applications, and while it is common to exclude cases with missing covariates, this approach can significantly bias resulting analyses. Missing data has received extensive attention in the statistical literature, while the EM-algorithm and multiple imputation are common approaches in practice \cite{little2019statistical}. In this paper, we explore an alternative Metropolis-Hastings algorithm, which has the following advantages over the aforementioned competitors. 

First, there are far fewer assumptions that are required to be able to implement this algorithm. For example, there are no assumptions that the data must be missing at random (MAR) or missing completely at random (MCAR), there are no assumptions that the data must possess a monotone missingness pattern, the distribution of the covariates does not have to possess a certain functional form, and the choice of the prior is unrestricted. 

Second, it provides joint inferences on all parameters. We group together the parameters of these models into three groups: the parameters for the conditional likelihood (these are the regression coefficients that are of primary interest to us, usually), the parameters of the distribution of the covariates that are missing, and the parameters that govern the missingess-mechanism. This procedure yields samples for all three groups.

Third, there is no logical inconsistency of assuming there are two different posterior distributions. Multiple imputation procedures aim to draw from the posterior predictive distribution of the model, but often do this in a way that does not take into account the specific structure of a given problem. This is often practical, because the imputation task and the complete-data modeling task can be done separately. With our approach, imputing and modeling can be done separately in some sense; however, only one algorithm brings to bear the effort of these two parties.

There is a substantial computational cost to gaining these benefits, however. For nontrivial data sets, every MCMC iteration might require a very large number of simulations. This difficulty necessitates careful programming, and puts limits on the kinds of data sets to which this methodology can be feasibly applied. We address this issue in later sections and detail some strategies that helped with our specific modeling tasks. While this method is developed fully in the context of logistic regression, extension to other models is immediate. Our work was motivated by an analysis of the field data risks and benefits associated with knee airbag deployment for occupants involved in frontal car crashes; a sub-analysis of this data is developed as an application.  

\section{Logistic Regression and Missing Data}

\subsection{Our Model's Assumptions}

Let $\mathbf{y} = (y_1, \ldots, y_n)^{\intercal}$ be the vector of completely observed binary responses for units $i=1,\ldots,n$, $\mathbf{x} = (\mathbf{x}_1, \ldots, \mathbf{x}_n)^{\intercal}$ the matrix of the covariate data where each $\mathbf{x}_i$ is the $p \times 1$ vector of covariate data for unit $i$, and let $\bm{\beta}$ be the vector of regression coefficients. 

Again, we assume that all response variables are observed, but that there are missing data in the covariates. Denote by $\text{mis}  = \{(i,j) : x_{i,j} \text{ is observed}   \}$ and $\text{obs}  = \{(i,j) : x_{i,j} \text{ is not observed}   \}$. Also denote  $\text{mis}_i  = \{ j : x_{i,j} \text{ is not observed}   \}$ and $\text{obs}_i  = \{j : x_{i,j} \text{ is observed} \}$. 

Having access to all of the covariate data would allow one to evaluate the complete conditional likelihood
\begin{gather} \label{eqn:cndtl_like}
p(\mathbf{y} \mid \mathbf{x},\bm{\beta}) = \prod_{i=1}^n \left(\frac{\exp(\mathbf{x}_i^{\intercal} \bm{\beta}) }{1 + \exp(\mathbf{x}_i^{\intercal} \bm{\beta})}\right) ^{y_i} \left(1-\frac{\exp(\mathbf{x}_i^{\intercal} \bm{\beta}) }{1 + \exp(\mathbf{x}_i^{\intercal} \bm{\beta})}\right)^{1-y_i}.
\end{gather}

We also specify the distribution of the covariate data that are missing. We assume that all rows are independent, and that for each row, the distribution of the unobserved data in column $j$, parameterized by a vector $\bm{\alpha}$, depends on columns $1$ through $j-1$:
\begin{gather} \label{eqn:missdata}
p(\mathbf{x}_{\text{mis}} \mid \mathbf{x}_{\text{obs}}, \bm{\alpha}) = \prod_{\{i : (i,j) \in \text{mis}\}} \prod_{j \in \text{mis}_i } f_{j}(\mathbf{x}_{i,j} \mid \mathbf{x}_{i,1:j-1}, \bm{\alpha}). 
\end{gather}
To simplify notation, we follow the convention where if $j=1$, $f_{1}(\mathbf{x}_{i,1} \mid \mathbf{x}_{i,1:1-1}, \bm{\alpha}) = f_{1}(\mathbf{x}_{i,1} \mid \bm{\alpha})$. This structure follows along the lines of \cite{mdmfglms}. Note that the distribution of a missing observation $\mathbf{x}_{i,j}$ may depend on both missing and observed data. 

Finally, we will make use of the {\it inclusion indicator}, an $n \times p$ matrix $\mathbf{m}$ whose $(i,j)$ element is $1$ if $x_{i,j}$ is observed, and $0$ otherwise. The missing-data mechanism is the conditional probability distribution of this indicator matrix:
\begin{gather}\label{eqn:pm}
p(\mathbf{m} \mid \mathbf{y}, \mathbf{x}_{\text{mis}}, \mathbf{x}_{\text{obs}}, \bm{\phi}).
\end{gather}

It is common to assume that this distribution is free of either the missing data, or free of both missing and observed data. These assumptions are termed missing at random and missing completely at random, and they are indispensable in showing that the missing-data mechanism is ignorable, or in other words, that samples from the posterior can be obtained ignoring missing values and the places in which those values are missing \cite{gelman2013bayesian}. There is no such assumption made in this paper, however, because this is not a requirement of the described methodology.

\section{Markov Chain Monte Carlo Approaches}

After we have chosen a prior distribution for our parameters, call it $p(\bm{\alpha}, \bm{\beta})$, we aim to draw samples from the marginal posterior $p(\bm{\beta} \mid \mathbf{y}, \mathbf{x}_{\text{obs}})$. This is the distribution of everything we are interested in, conditioning on everything we know. Because we are interested in being able to sample from this distribution without restricting any choice of a prior distribution or a missing-data mechanism, we will use a variant of the Metropolis-Hastings algorithm \cite{mh}, \cite{gelman2013bayesian}. This article will accomplish this by taking the approach of sampling from a joint posterior defined on a space of much larger dimension, and integrating out nuisance parameters. 

\subsection{Initial Approaches}

First, one may look at the high-dimensional posterior of everything that is unknown: $p(\bm{\alpha}, \bm{\beta}, \bm{\phi}, \mathbf{x}_{\text{mis}} \mid \mathbf{y}, \mathbf{x}_{\text{obs}}, \mathbf{m})$, which is proportional to
\[
p(\mathbf{m} \mid \mathbf{y}, \mathbf{x}_{\text{mis}}, \mathbf{x}_{\text{obs}}, \bm{\phi})p(\mathbf{y} \mid \mathbf{x}_{\text{mis}}, \mathbf{x}_{\text{obs}}, \bm{\beta}) p(\mathbf{x}_{\text{mis}} \mid \mathbf{x}_{\text{obs}}, \bm{\alpha}) p(\bm{\alpha}, \bm{\beta}, \bm{\phi}).
\]
It is {\it theoretically} possible to use the Metropolis-Hastings algorithm to sample the parameters as well as the missing data all at once, because this function above is proportional to the target density; however, {\it practically} speaking, this approach might be infeasible if there are too many missing data points.   

A second approach would avoid sampling on this unnecessarily large space by using the ``marginal" MH algorithm, which targets
\begin{align}\label{eqn:obs_data_like}
p(\bm{\alpha}, \bm{\beta}, \bm{\phi} \mid \mathbf{y}, \mathbf{x}_{\text{obs}}, \mathbf{m}) &\propto p(\mathbf{m}, \mathbf{y} \mid \mathbf{x}_{\text{obs}}, \bm{\alpha}, \bm{\beta}, \bm{\phi}) p(\bm{\alpha}, \bm{\beta}, \bm{\phi}).
\end{align}
This approach would solve the problem of dimensionality if it was possible to evaluate the ``observed-data likelihood" in the right hand side of Equation \ref{eqn:obs_data_like}. Indeed, if one used the Metropolis-Hastings algorithm \cite{mh}, one would simulate a Markov chain $\{\bm{\alpha}^k, \bm{\beta}^k, \bm{\phi}^k \}_k$ which leaves the normalized version of that expression invariant. Given that we are at $\bm{\alpha}, \bm{\beta}, \bm{\phi}$ in the chain, we first propose $(\bm{\alpha}', \bm{\beta}', \bm{\phi}') \mid (\bm{\alpha}, \bm{\beta}, \bm{\phi}) \sim q_{\text{MH} }(\cdot \mid \bm{\alpha}, \bm{\beta}, \bm{\phi})$, and accept this proposal with probability  
\begin{eqnarray} \label{eqn:ideal_accept_prob}
\min\left[ 1, \frac{ p(\mathbf{m}, \mathbf{y} \mid \mathbf{x}_{\text{obs}}, \bm{\alpha}', \bm{\beta}', \bm{\phi}')  p(\bm{\alpha}', \bm{\beta}', \bm{\phi}') q_{\text{MH} }(\bm{\alpha}, \bm{\beta}, \bm{\phi} \mid \bm{\alpha}', \bm{\beta}', \bm{\phi}') }{ p(\mathbf{m}, \mathbf{y} \mid \mathbf{x}_{\text{obs}}, \bm{\alpha}, \bm{\beta}, \bm{\phi})  p(\bm{\alpha}, \bm{\beta}, \bm{\phi}) q_{\text{MH} }(\bm{\alpha}', \bm{\beta}', \bm{\phi}' \mid \bm{\alpha}, \bm{\beta}, \bm{\phi})}\right].
\end{eqnarray}

Typically the dimensions of $\bm{\alpha}$, $\bm{\beta}$, and $\bm{\phi}$ are small or moderate, so this would be a relatively ideal situation. However, it is rare to be able to evaluate the observed-data likelihood because it is often a high-dimensional integral, and one would rather not make overly-restrictive assumptions about the missing-data mechanism \cite{mdmfglms}.

\subsection{A Pseudo-Marginal Metropolis-Hastings Algorithm}

The pseudo-marginal algorithm \cite{Beaumont}, \cite{andrieu2009} resembles the above algorithm; however, it replaces the intractable quantity with an estimate that is nonnegative and unbiased. The algorithm described above remains the same, with the exception that the acceptance probability at some iteration becomes
\begin{eqnarray} \label{eqn:accept_prob}
\min\left[ 1, \frac{\hat{p}(\mathbf{m}, \mathbf{y} \mid \mathbf{x}_{\text{obs}}, \bm{\alpha}', \bm{\beta}', \bm{\phi}')p(\bm{\alpha}', \bm{\beta}', \bm{\phi}') q_{\text{MH}}(\bm{\alpha}, \bm{\beta}, \bm{\phi} \mid \bm{\alpha}', \bm{\beta}', \bm{\phi}') }{\hat{p}(\mathbf{m}, \mathbf{y} \mid  \mathbf{x}_{\text{obs}}, \bm{\alpha}, \bm{\beta}, \bm{\phi})p(\bm{\alpha}, \bm{\beta}, \bm{\phi})q_{\text{MH}}(\bm{\alpha}', \bm{\beta}', \bm{\phi}'\mid \bm{\alpha}, \bm{\beta}, \bm{\phi})}\right],
\end{eqnarray}
where $\hat{p}(\mathbf{m}, \mathbf{y} \mid \mathbf{x}_{\text{obs}}, \bm{\alpha}, \bm{\beta}, \bm{\phi})$ is an unbiased estimate of $p(\mathbf{m}, \mathbf{y} \mid \mathbf{x}_{\text{obs}}, \bm{\alpha}, \bm{\beta}, \bm{\phi})$ Despite (4) not being equal to (5), the sampler targets the left hand side of (4) exactly. 

With this particular model, we estimate the likelihood at each iteration using importance sampling \cite{kahnharris}. First, choose an importance distribution for the missing data, call it $q_{\text{IS}}( \mathbf{x}_{\text{mis}} \mid \mathbf{x}_{\text{obs}}, \mathbf{y})$. This distribution must be positive whenever $p(\mathbf{m} \mid \mathbf{y}, \mathbf{x}_{\text{mis}}, \mathbf{x}_{\text{obs}}, \bm{\phi}) p(\mathbf{y} \mid \mathbf{x}_{\text{mis}}, \mathbf{x}_{\text{obs}}, \bm{\beta}) p(\mathbf{x}_{\text{mis}} \mid \mathbf{x}_{\text{obs}}, \bm{\alpha})$ is also positive. Then choose a number of samples $N$, and for $k=1, \ldots N$, draw $\mathbf{x}^k_{\text{mis}} \sim q_{\text{IS}}( \mathbf{x}_{\text{mis}} \mid \mathbf{x}_{\text{obs}}, \mathbf{y})$ and evaluate
\begin{gather} \label{eqn:impsamp}
\hat{p}(\mathbf{m}, \mathbf{y} \mid \mathbf{x}_{\text{obs}},\bm{\alpha}, \bm{\beta}, \bm{\phi}) = \\ \nonumber
N^{-1} \sum_{k = 1}^{N}  \frac{p(\mathbf{m} \mid \mathbf{y}, \mathbf{x}^k_{\text{mis}}, \mathbf{x}_{\text{obs}}, \bm{\phi}) p(\mathbf{y} \mid \mathbf{x}^k_{\text{mis}}, \mathbf{x}_{\text{obs}}, \bm{\beta}) p(\mathbf{x}_{\text{mis}}^k \mid \mathbf{x}_{\text{obs}}, \bm{\alpha}) }{ q_{\text{IS}}( \mathbf{x}^k_{\text{mis}} \mid \mathbf{x}_{\text{obs}}, \mathbf{y}) } .
\end{gather}
By the strong law of large numbers, this converges to the true marginal likelihood as $N \to \infty$. More importantly, it is unbiased, which is the primary requirement of the pseudo-marginal approach. The entire algorithm is described in Algorithm \ref{alg:PMMH}.

\begin{algorithm}
\centering
\caption{a PMMH algorithm}\label{alg:PMMH}
{\fontsize{8}{2}\selectfont
\begin{algorithmic}[0]
\Procedure{pmmh}{$N, q_{\text{MC}}, q_{\text{IS}} $}
  \If{$i$ equals $1$}
    \State choose and store $\bm{\alpha}^1, \bm{\beta}^1, \bm{\phi}^1$
    \State store $\hat{p}(\mathbf{m}, \mathbf{y} \mid \mathbf{x}_{\text{obs}},\bm{\alpha}^1, \bm{\beta}^1, \bm{\phi}^1)$
  \Else
    \State draw $\bm{\alpha}^{\prime}, \bm{\beta}^{\prime}, \bm{\phi}^{\prime} \sim q_{\text{MH}}(\cdot \mid \bm{\alpha}^{i-1}, \bm{\beta}^{i-1}, \bm{\phi}^{i-1})$
	\State calculate and store $\hat{p}(\mathbf{m}, \mathbf{y} \mid \mathbf{x}_{\text{obs}}, \bm{\alpha}^{\prime}, \bm{\beta}^{\prime}, \bm{\phi}^{\prime})$
	\State draw $U \sim \text{Uniform}(0,1]$
    \If{$U < \min\left[ 1, \frac{\hat{p}(\mathbf{m}, \mathbf{y} \mid \mathbf{x}_{\text{obs}}, \bm{\alpha}', \bm{\beta}', \bm{\phi}')p(\bm{\alpha}', \bm{\beta}', \bm{\phi}') q_{\text{MH}}(\bm{\alpha}^{i-1}, \bm{\beta}^{i-1}, \bm{\phi}^{i-1} \mid \bm{\alpha}', \bm{\beta}', \bm{\phi}') }{\hat{p}(\mathbf{m}, \mathbf{y} \mid  \mathbf{x}_{\text{obs}}, \bm{\alpha}^{i-1}, \bm{\beta}^{i-1}, \bm{\phi}^{i-1})p(\bm{\alpha}^{i-1}, \bm{\beta}^{i-1}, \bm{\phi}^{i-1})q_{\text{MH}}(\bm{\alpha}', \bm{\beta}', \bm{\phi}'\mid \bm{\alpha}^{i-1}, \bm{\beta}^{i-1}, \bm{\phi}^{i-1})}\right]$}
\State $\bm{\alpha}^i, \bm{\beta}^i, \bm{\phi}^i \gets \bm{\alpha}^{\prime}, \bm{\beta}^{\prime}, \bm{\phi}^{\prime}$
    \Else
      \State $\bm{\alpha}^i, \bm{\beta}^i, \bm{\phi}^i \gets \bm{\alpha}^{i-1}, \bm{\beta}^{i-1},\bm{\phi}^{i-1}$
    \EndIf
  \EndIf
\EndProcedure
\end{algorithmic}
}
\end{algorithm}

The question of how to choose the three tuning elements of this algorithm ($q_{\text{MC}}, q_{\text{IS}}, N$) is still an open one. When choosing $q_{\text{MH}}$ for use in a more classical algorithm, wherein likelihood evaluations are available, there are recommendations for target acceptance rates, as well as recommendations for how to choose the proposal's parameters \cite{gelman2013bayesian}. Regarding the importance sampling importance distribution, $q_{\text{IS}}$, the asymptotic variance of its estimator, when it exists, will be large when the ratio between the importance distribution's density and the density of the target is far from unity for $q_{\text{IS}}$-likely values. In particular, it is problematic when the importance density has thinner tails than the target, when the target is relatively peaked, or when the target is defined on a high-dimensional space \cite{mcbook}. The last difficulty is especially problematic for this technique. For a fixed percentage of the data that is missing, there will always be a sufficiently large size of data that renders this approach infeasible. 

The dispersion of the estimates of the log-likelihood can often be the deciding factor when it comes to determining whether a chain mixes well or not. \cite{andrieu2016} show that increasing $N$ not only decreases the variance of the importance sampling estimate, but that it lowers the asymptotic variance of parmeter estimates taken from the pseudo-marginal chain. However, there are computational limits on how far $N$ can be increased. Recently, much work has provided guidelines for choosing $N$. \cite{pmmh_tuning} and \cite{sherlock2015} describe guidelines for choosing $N$ in order to minimize computing time criteria. These papers both suggest choosing a value of $N$ that makes the variance of the log-likelihood between $1$ and $2$.

Increasing the precision of your likelihood estimates without resorting to increasing $N$ is obviously extremely useful as well; here knowledge of classical variance-reduction techniques is indispensable. \cite{corrpm} describe how to correlate the estimates of the log-likelihood at consecutive iterations in order to minimize the variance of their difference. 

In practice, tuning the estimation algorithm starts with choosing an importance sampling strategy first, then choosing $N$ second. The importance sampling strategy will often be close to the best a user can come up with. Afterwards, several $N$s are tried. One might compute many log-likelihood estimates, and increase $N$ until the sample variance is below $2$, say. This is far more convenient than running ``full-length" simulations that are increasingly more computationally expensive; however, for nontrivial problems, this decision rule might suggest values of $N$ that are too high to be practically useful. For example, if this strategy suggests an $N$ that is so large that one would only be able to come up with a few hundred samples, it's not clear that one should follow this rule as it is unlikely that the chain will have enough time to burn-in or converge to stationarity. 

TODO: edit this The current state is far from totally determined, and because the specifics of our problem do not satisfy all of the assumptions used in the above-mentioned work, we choose to take a more applied approach. We focus on choosing $q_{\text{IS}}$, and then select the other tuning parameters based on several ``pilot runs" of the algorithm. We also follow along the lines of (don't cite this) and parallelize the computation of each iteration's importance sampling estimator.

\subsection{A Comparison with Multiple Imputation}

Multiple Imputation (MI) generates multiple complete data sets by sampling several sets of plausible values for each missing data point by sampling from the posterior predictive distribution \cite{rubin:1978}, \cite{rubin:1987}, \cite{gelman2013bayesian}. The same analysis is performed separately on each  data set, and the results are then combined. For example, in the context of regression analysis, the model parameters derived from each imputed dataset are combined by a simple average. The parameter variances are calculated by averaging the individual variances from each imputation, and the formula includes an additional term to capture the between-imputation variance.

\cite{carlin_lee} compare the two implementations most commonly used in practice: Multivariate Normal Imputation (MVNI) and Multivariate Imputation by Chained Equations (MICE). MVNI was proposed by \cite{Schafer:1997}, and is implemented in software packages such as \cite{amelia}. It assumes the data are normally distributed, and that the missing values are missing at random. MICE, on the other hand, also known as the Full Conditional Specification, is more flexible. It does not assume that the data are normally distributed, and so it is more capable of handling binary and ordinal data. Several statistical software packages implement this procedure \cite{mi}, \cite{mice}. First, initial estimates for the missing values are drawn from the existing data, and then, columns of data are sampled sequentially. If the missing pattern is monotone, one sweep through all of the column is required. Otherwise, Gibbs sampling is performed, which alternates between draws of the conditional distributions until a convergence criterion is reached \cite{Raghunathan01amultivariate}.

MI is popular because it divides the labor of data analysis between the data collector and the data analyst. The data collector may impute multiple data sets on his own, taking into account his knowledge of the data collection process, and then send these on to the data analyst, who performs the same analysis on each imputed data set \cite{little2019statistical}. 

However, despite its convenience, MI has some obvious drawbacks. First, it is logically inconsistent to assume that there are two posterior distributions, one used for imputing data, and one used for performing the final analysis. It is hard to say how this affects how inaccurate the resulting estimates will be. Second, the algorithm is potentially more inefficient. If one is already drawing parameters along with missing data values, why throw away those parameters before drawing them again? Currently, the second approach is not likely to ``stick," as the pseudo-marginal method is notoriously computationally expensive, as we will see in the applied sections that follow.

\section{A Simulation Study}

We perform a simulation study to demonstrate the algorithm's ability to recover the true parameters of the model specified by Equations \ref{eqn:cndtl_like}, \ref{eqn:missdata}, and \ref{eqn:pm}. We show the performance of the algorithm for different tuning parameters, and discuss some common difficulties of implementation. All of the code is provided by the authors; more details can be found in the appendix.

\subsection{The Data}

Our simulated data set has two covariate columns: $\mathbf{x}_1$ and $\mathbf{x}_2$. The first is fully-observed, and the second has 34\% of its $100$ rows missing. To generate the data, we assume $x_{i1} \overset{iid}{\sim} \text{Normal}(0,1)$, and $x_{i2} \overset{iid}{\sim} \text{Normal}(0,\alpha)$.
 So, in particular, if $x_{i,2}$ is missing, 
$$
p(x_{\text{mis}_i} \mid x_{\text{obs}_i}, \bm{\alpha}) = p(x_{i2} \mid \bm{\alpha}) = \text{Normal}(0, \alpha).
$$	
If $x_{i2}$ is not missing, then the left hand side of the above display is $1$.

We keep the conditional likelihood as Equation \ref{eqn:cndtl_like}, and we use a similar form to specialize Equation \ref{eqn:pm}. More specifically, we assume the missing-data mechanism is

$$
p(\mathbf{m} \mid \mathbf{y}, \mathbf{x}_{\text{mis}}, \mathbf{x}_{\text{obs}}, \bm{\phi}) = \prod_{i=1}^n \prod_{j=1}^2 p(m_{ij} \mid y_i, x_{i1}, x_{i2}, \bm{\phi}) 
$$
where $p(m_{ij}=1 \mid y_i, x_{i1}, x_{i2}, \bm{\phi})$ equals
$$
[\text{IL}(\phi_0 + \phi_1 x_{i1} + \phi_2 x_{i2} )]^{m_i}[1-\text{IL}(\phi_0 + \phi_1 x_{i1} + \phi_2 x_{i2} )]^{1-m_i} 
$$
if $j=2$, otherwise the left hand side of the above display is $1$. Here we are using $\text{IL}$ to refer to the ``inverse logit" or ``logistic sigmoid" function.

Familiar distributions are chosen for the priors. We let $p(\alpha)$ be a $\text{Inverse Gamma}(1.65, .65)$ distribution, $p(\bm{\beta})$ is chosen to be a $\text{Normal}(\mathbf{0}, 3 \mathbf{I})$, and $p(\bm{\phi})$ is made to be a $\text{Normal}(\mathbf{0}, 3\mathbf{I} )$. These priors were selected by a hypothetical analyst, and were not used to simulate the the data-generating parameters. Rather, the ``true" data-generating parameters are $\alpha_{\text{true}} = 1$, $\bm{\beta}_{\text{true}} = [1, -2, 3]^\intercal$, and $\bm{\phi}_{\text{true}} = [1,1,1]^\intercal$.

\subsection{The PMMH Approach}

There are two proposal distributions that we must choose: the importance sampling proposal distribution $q_{\text{IS}}$, and the Metropolis-Hastings proposal distribution $q_{\text{MH}}$. For the latter, we transform the parameters to be real-valued, and then propose new values using a random walk on this transformed space. The regression coefficients in the conditional likelihood and the missing-data mechanism are already real-valued, so only $\alpha$ will be transformed. It will be transformed into $\log (\alpha)$, where $\log$ denotes the base $e$ logarithm.

Regarding the importance sampling proposal distribution, we choose 
$$
q_{\text{IS}}( \mathbf{x}_{\text{mis}} \mid \mathbf{x}_{\text{obs}}, \mathbf{y}) = \prod_{\{i : (i,j) \in \text{mis} \}}^n q_{\text{IS}}(x_{i2} \mid x_{i1}),
$$ where $q_{\text{IS}}(x_{i2} \mid x_{i1}) = t_{10}\left(0, \alpha \right)$, which is a scaled $t$-distribution with $10$ degrees of freedom.

$100,000$ iterations were performed, and at each iteration, Equation \ref{eqn:accept_prob} is evaluated using $N=5000$. After discarding $100$ samples for burn-in, estimates, $95$\% credible intervals, and $\hat{R}$ convergence diagnostics \cite{gelman2013bayesian} are calculated for every element of $(\alpha, \boldsymbol{\beta}^\intercal, \boldsymbol{\phi}^T)$ (this was done by splitting the single chain into two parts after discarding $100$ iterations as burn-in). This information is provided in Table \ref{tab:fake_data_results}. Histograms of each parameter's samples are provided in the appendix in Figure \ref{fig:fake_data_samples}. Do note that the samples for $\phi_2$ are not completely satisfactory. This is reflected in the histogram spike at $0$, and in the $\hat{R}$ value that is slightly high.

\begin{table}[ht]
\centering
\begin{tabular}{rcclll}
  \hline
  param. & truth & estimate & $95$\% cred. & $\hat{R}$ \\ 
  \hline
  $\alpha$ & 1 & 1.43 & (0.31, 3.78) & 1.02 \\ 
  $\beta_0$ & 1 & 1.35 & (0.91, 1.78) & 1.01 \\ 
  $\beta_1$ & -2 & -2.44 & (-3.94, -1.35) & 1.02 \\ 
  $\beta_2$ & 3 & 3.88 & (2.35, 5.58) & 1.08 \\ 
  $\phi_0$ & 1 & 1.06 & (0.51, 1.8) & 1.15 \\ 
  $\phi_1$ & 1 & 1.08 & (0.4, 1.74) & 1.01 \\ 
  $\phi_2$ & 1 & 0.76 & (-0.15, 1.38) & 1.29 \\   
   \hline
\end{tabular}
\caption{Pseudo-Marginal MH results on our simulated data}\label{tab:fake_data_results}
\end{table}

\subsection{Using MICE}

The same data was analyzed using MICE. We made use of the \verb|mice| package in \verb|R| \cite{mice}. Ten data sets were imputed using the method of Bayesian linear regression, using both $\mathbf{y}$ and $\mathbf{x}_1$ values as predictors. Only estimates for the regression coefficients are available. Improvized confidence intervals are created by adding and subtracting $1.96$ standard errors from each parameter estimate. This information is given below in Table \ref{tab:fake_data_results2}.

\begin{table}[ht]
\centering
\begin{tabular}{rccl}
  \hline
 param. & truth & estimate & $95$\% interval \\ 
  \hline
  $\beta_0$ & 1 & 0.95 & (-0.11, 2.01) \\ 
  $\beta_1$ & -2 & -2.00 & (-3.58, -0.42) \\ 
  $\beta_2$ &  3 & 3.40 & (0.96, 5.83) \\ 
   \hline
\end{tabular}
\caption{MICE results on our simulated data}\label{tab:fake_data_results2}
\end{table}

It is problematic to compare credible intervals with confidence intervals, but a few things are worth pointing out. One will notice that the PMMH method yields narrower intervals. Moreover, these intervals were constructed in a less ambiguous way. They are calculated by taking empirical quantiles of the samples, so they take into account the shape of each marginal posterior better than adding and subtracting a single quantity from a point estimate. MICE produced point estimates that happen to be closer to the data-generating values. This could be just a coincidence. After all, MICE did not take into account our known missingness-mechanism. 

Regarding the cost of computation, it is unfair to compare packaged code with a pure \verb|R| implementation, so we do not provide any runtime measurements. Two things are certain, though. First, PMMH is much more computationally demanding than MICE, and it is unlikely that further refining the PMMH code would close this time gap completely. This particular script takes around $24$ hours to run, whereas with MICE, it takes just a brief moment. Second, speeding up PMMH in \verb|R| would be a very fruitful computational undertaking. Compiled languages would be useful for their overall quickness. In particular, it would be useful to expose a framework that uses pass-by-reference semantics to facilitate the use of common random numbers. After this has been completed, more involved simulation experiments would be very informative.

The critical aspect of the pseudo-marginal algorithm is the dispersion of the quantity in Equation \ref{eqn:impsamp}. We demonstrate this with the following plots, which are something akin to a ``profile" surface. The calculation of $\hat{p}(\mathbf{m}, \mathbf{y} \mid \mathbf{x}_{\text{obs}},\bm{\alpha}, \bm{\beta}, \bm{\phi})$ used in Algorithm \ref{alg:PMMH} takes as inputs of all the parameters, as well as $N$ replications of $x_{\text{mis}}$ (despite its notation not clearly reflecting that). In an effort to plot $(\alpha, \beta_2) \mapsto - \log \hat{p}(\mathbf{m}, \mathbf{y} \mid \mathbf{x}_{\text{obs}},\bm{\alpha}, \bm{\beta}, \bm{\phi})$, we set all other parameters equal to the data-generating values, and we simulate $N = 2, 5, 15$ missing data sets. For each sample size, two surface plots are provided, each of which use different replications of $x_{\text{mis}}$. 

\begin{figure}
    \centering
    \begin{subfigure}[b]{0.31\textwidth}
        \includegraphics[width=\textwidth]{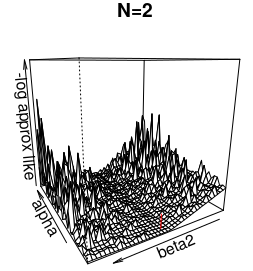}
        \label{fig:N21}
    \end{subfigure}
    ~ 
    \begin{subfigure}[b]{0.31\textwidth}
        \includegraphics[width=\textwidth]{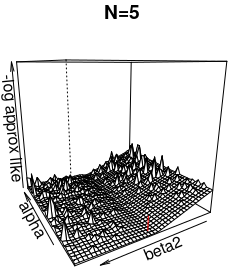}
        \label{fig:N51}
    \end{subfigure}
    ~ 
    \begin{subfigure}[b]{0.31\textwidth}
        \includegraphics[width=\textwidth]{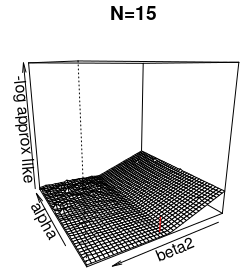}
        \label{fig:N151}
    \end{subfigure}

    \begin{subfigure}[b]{0.31\textwidth}
        \includegraphics[width=\textwidth]{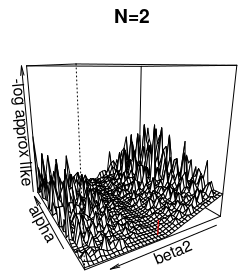}
        \label{fig:N21}
    \end{subfigure}
    ~ 
    \begin{subfigure}[b]{0.3\textwidth}
        \includegraphics[width=\textwidth]{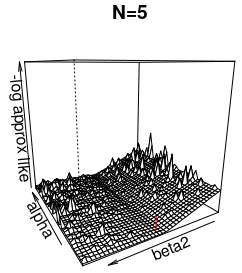}
        \label{fig:N51}
    \end{subfigure}
    ~ 
    \begin{subfigure}[b]{0.3\textwidth}
        \includegraphics[width=\textwidth]{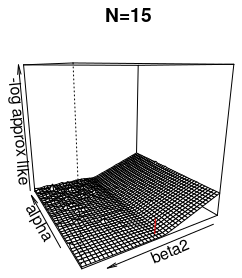}
        \label{fig:N151}
    \end{subfigure}
    
    \caption{An approximation of $(\alpha, \beta_2) \mapsto - \log \sup_{\beta_0, \beta_1, \bm{\phi}} \hat{p}(\mathbf{m}, \mathbf{y} | \mathbf{x}_{\text{obs}},\bm{\alpha}, \bm{\beta}, \bm{\phi})$}\label{fig:rough_surfaces}
\end{figure}

Most noticeably, a lower number of importance samples $N$ produces a rougher surface. This means that, even if a ``likely" parameter vector is proposed by $q_{\text{MH}}$, it could still fall on a spike and thus be rejected with virtual certainty. Moreover, it is hard to control the location of these spikes. When comparing the plots in a given column, it can be seen that the location of these spikes changes, and thus only depends on the missing data that is simulated {\it after} parameters have been proposed. 

It is unsurprising that $\alpha$ seems to be less identified than $\beta_2$, but it is interesting to note that spikes are less severe near values of $\alpha=0$. When $\alpha$ is near $0$, the importance weights found in (\ref{eqn:impsamp}) are nearly constant for most sampled data values. In short, approximating the acceptance probability is much easier for certain areas of the parameter space. Taken together, these plots suggest that raising the sample size will increase smoothness more in some areas and less in others, and that the acceptance rate of this algorithm will not necessarily be improved by tuning the MH proposal $q_{\text{MH}}(\cdot \mid \theta)$.

\section{Analysis of a Real Data Set}

\subsection{Background}


This work was originally motivated by an analysis of data collected by the National Automotive Sampling System, Crashworthiness Data System (NASS-CDS) to understand the potential benefits and risks associated with knee airbag (KAB) deployment for belted occupants in real-world frontal car crashes. NASS-CDS was a survey sample of approximately 5,000 tow-away crashes in the United States each year, conducted by the National Highway Traffic Safety Administration (NHTSA) \cite{CDS_site}. Each sampled crash received a thorough investigation which included crash reconstruction, demographics of all involved occupants, vehicle characteristics (year, make, model, etc) of all involved vehicles, thorough documentation of vehicle damage including photos, and documentation of all occupant injuries using the Abbreviated Injury Scale (AIS) \cite{Gennarelli2006}.

Injury risk in frontal crashes depends significantly on the vehicle's change in velocity (delta-V), which is normalized to be ``barrier impact equivalent" and is estimated as part of the crash reconstruction. Additional significant predictors are occupant age and sex. Vehicles tend to get modestly safer with each additional model year. Factors such as direction of force (measured in in 10$^\circ$ increments), occupant height and weight (or body mass index), and vehicle type are considered to be important covariates, but are not consistently associated with increases or decreases in risk in field data analyses of frontal crashes \cite{doi:10.1080/15389588.2018.1528356}.

Missing data is a significant problem for all analyses using NASS-CDS. Approximately 30\% of all frontal crashes do not have an estimated delta-V, and a different 30\% are missing occupant height and weight. The data also contains auxiliary information, such as vehicle crush measurements, which in some cases can inform imputation. Data does not appear to be missing completely at random. For example, crashes with missing delta-V seem to have a higher proportion of uninjured occupants. Nonetheless, it is desirable to include all cases in order to both minimize bias from excluded cases and to retain all occupants who experienced KAB deployment; KABs have recently become more common in the vehicle fleet, but at this point, only relatively few crashes of KAB equipped vehicles have been investigated. 

For exposition, in this article, we investigate the probability of the occupant receiving a serious or worse injury (AIS 3 or higher) in any body region. The primary predictor of interest is knee airbag deployment, and we wish to adjust for the covariates indicated above while including well-established prior information about effect sizes, as well as provide realistic models for the missing covariates. 

\subsection{Results}

The following variables are used in our model as predictors: age (in years), sex, body mass index (BMI), the sum of two delta-V measurements (\verb|dvtotal|), whether the knee airbag was deployed, whether the vehicle was a sport utility vehicle (SUV), whether the vehicle was a truck, whether the vehicle was a van, model year (centered at 2003), and a categorical direction of force variable. Among these, the variables that exhibit some missingness are sex, BMI, \verb|dvtotal|, and model year (see Table \ref{tab:perc_missing}). To help with sampling missing data, we make use of a variable \verb|dvc|, which is a sum of several measurements of how much a vehicle was crushed during impact. Interestingly, some of this column is missing as well, so it too must have its distribution specified.  

\begin{table}[ht]
\centering
\begin{tabular}{rr}
  \hline
  variable & \% miss \\ 
  \hline
  \verb|sex| & 0.030 \\ 
  \verb|bmi| & 13.050 \\ 
  \verb|dvtotal| & 29.023 \\ 
  \verb|modelyr| & 0.004 \\ 
  \verb|dvc| & 20.241 \\ 
  \hline
\end{tabular}
\caption{Perent missingness of covariates.} \label{tab:perc_missing}
\end{table}

To specify the covariates' distributions and the parameter priors, we make use of considerable subject-matter expertise. These 23 distributions are listed in Appendix \ref{appendix}, and are briefly described here. All of the regression coefficients were given normal priors, and the parameters of these distributions were carefully chosen to cohere with the knowledge obtained from similar studies in the past. The missing data distributions were chosen so that they would generate data that looks similar to what is observed. In particular, the empirical support of each distribution was first noted. For example, the dummy variable for sex was given a Bernoulli distribution, and the missing crush measurements, because their observed values only took on integers, were given a Negative Binomial distribution. The data for BMI was positively skewed, so we gave its missing data a Skew Normal distribution. The priors were informative wherever appropriate. Location-scale families were elicited directly, while Beta and Inverse-Gamma distributions were elicited from their prior means and quantiles. 

We also chose five distributions to help propose missing covariates to be used in calculating our importance sampling estimates. We generally tried to choose these distributions so that they would be visually similar to the empirical distributions of each covariate. Wherever possible, we give these distributions tails that are fatter than the missing data distributions, which is in accordance with a general principle of importance sampling. For example, BMI and model year were given Scaled-$t$ distributions. For more details, see Appendix \ref{appendix}. 

Only $15,000$ iterations were performed, each using an importance sampler with $N=5000$ samples. The program took slightly less than six days on a machine with an Intel® Core™ i7-4770 CPU @ 3.40GHz × 8 processor. The program was written entirely in \verb|R|, and, using the \verb|parallel| package \cite{par_package}, parallelization was employed to speed up the evaluation of the importance sampling estimator at each iteration.

Table \ref{tab:beta_table} shows the results for estimating $\bm{\beta}$ after discarding a thousand samples for burn-in. For comparison, the table provides the parameter estimates using a standard maximum likelihood estimation routine under the complete case analysis. The two procedures seem to yield estimates that are similar for the most part; however, there are some interesting differences to take note of. For example, our method's coefficient estimate for the dummy variable indicating whether the knee airbag was deployed has the opposite sign. Second, the magnitude of the effect size for sex seems to have increased. 

These results should be taken with a grain of salt, however, because the $\hat{R}$ diagnostics are relatively high. The interpretation of this diagnostic is the factor by which the scale of the current distribution for a particular univariate parameter ``might be reduced if the simulations were continued in the limit $n \to \infty$" \cite{gelman2013bayesian}. A common rule of thumb is to continue simulations until all of these values are less than $1.1$. More iterations could have been obtained if it were not for the expense of the computations. Clearly, there is a need for more work on the computations to be done here. Table \ref{tab:alpha_phi_table} in the appendix shows the parameter estimates for $\bm{\alpha}$ and $\phi$, along with the same uncertainty quantification used before. 

\begin{table}[ht]
\begin{tabular}{lllllll}
  \hline
 & Est. & MCSE & 95.cred.lower & 95.cred.upper & RHat & MLE\\ 
  \hline
  \verb|intercept| & -9.175 & 0.026 & -9.716 & -8.651 & 1.768 & -7.592 \\ 
  \verb|age| & 0.054 & 0.002 & -0.364 & 0.161 & 1.069 & 0.040 \\ 
  \verb|sex| & -2.963 & 0.023 & -3.263 & -1.372 & 1.209 & -0.366 \\ 
  \verb|bmi| & 0.054 & 0.004 & -0.233 & 0.201 & 1.077 & 0.019 \\ 
  \verb|dvtotal| & 0.102 & 0.004 & -0.166 & 0.351 & 1.193 & 0.082 \\ 
  \verb|kabdeply| & 0.891 & 0.048 & 0.342 & 2.230 & 1.538 & -0.692 \\ 
  \verb|suv| & -0.291 & 0.014 & -0.495 & 1.087 & 1.013 & -0.033 \\ 
  \verb|truck| & -1.685 & 0.013 & -2.049 & -1.342 & 1.127 & -0.274 \\ 
  \verb|van| & -0.180 & 0.012 & -1.214 & 0.173 & 1.003 & -0.136 \\ 
  \verb|modelyr| & -0.129 & 0.027 & -1.753 & 0.147 & 1.115 & -0.051 \\ 
  \verb|pdofgrFar| & -2.147 & 0.023 & -2.611 & -1.603 & 1.627 & -0.098 \\ 
  \verb|pdofgrNear| & 1.758 & 0.039 & 0.493 & 2.301 & 1.969 & 0.277 \\ 
   \hline
\end{tabular}
\caption{Regression coefficient estimates along with Monte Carlo Standard Error Estimates, 95\% credible intervals, Gelman-Rubin statistics. For comparison, complete case analysis maximum likelihood estimates are also provided in the last column} \label{tab:beta_table}
\end{table}

\section{Conclusion}

This paper has demonstrated the use of the pseudo-marginal Metropolis-Hastings algorithm to estimate generalized linear models in the presence of missing data in a fully Bayesian way, and it has mentioned some of the benefits this procedure possesses when being compared to other commonly used methods. However, this paper has also demonstrated that the feasibility of this approach is diminished when it is employed on data sets with many missing values, and so there is much work to be done to reduce the computational cost of this class of techniques. In our opinion, computational complexity is the primary impediment, and if future work reduces the computation time that is required, this approach could prove to be a promising tool that would enjoy a more widespread adoption.

\appendix

\section{Appendix}\label{appendix}

The appendix is divided into two sections: one for the results of the simulation study, and one for the results of the real-data analysis.

\subsection{Simulation Study}

\begin{figure}[H]
    \centering
    \begin{subfigure}[b]{0.41\textwidth}
        \includegraphics[width=\textwidth]{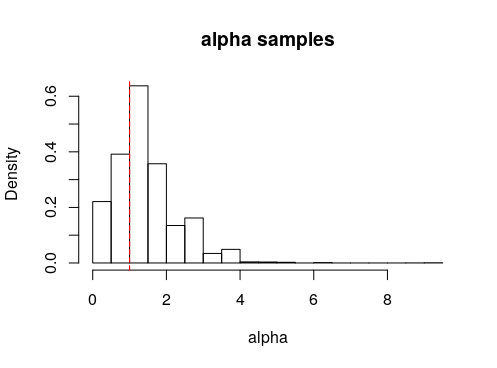}
        \label{fig:N21}
    \end{subfigure}

    \begin{subfigure}[b]{0.41\textwidth}
        \includegraphics[width=\textwidth]{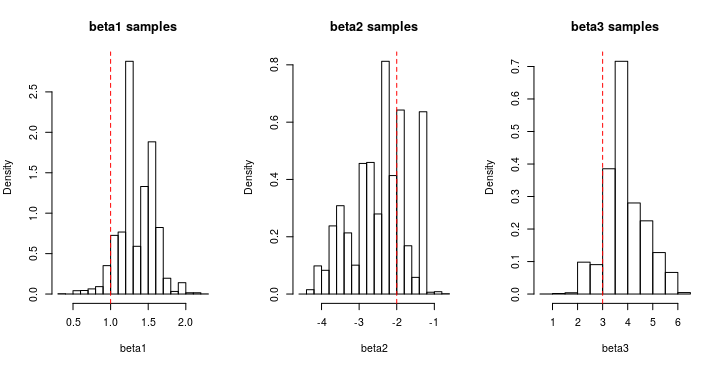}
        \label{fig:N51}
    \end{subfigure}

    \begin{subfigure}[b]{0.41\textwidth}
        \includegraphics[width=\textwidth]{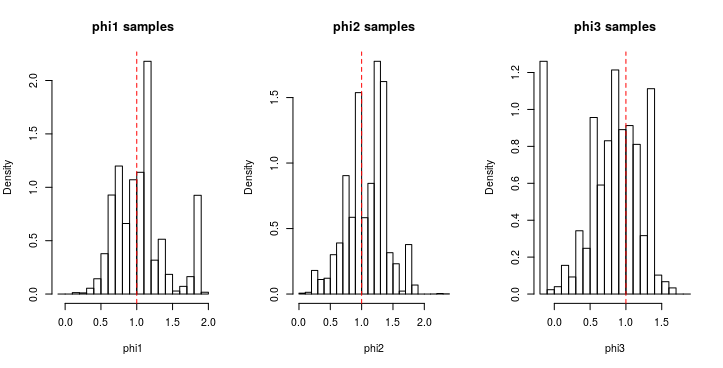}
        \label{fig:N151}
    \end{subfigure}
        
    \caption{Pseudo-Marginal Metropolis-Hastings samples using the simulated data set}\label{fig:fake_data_samples}
\end{figure}

\subsection{Real-Data Analysis}

Recall that the columns with missing values are sex, body mass index, dvtotal, and model year. DVC has missing data as well. Even though it's not a predictor for crash outcome, it must be imputed because the imputations of other variables rely on it. Referring to Equation \ref{eqn:missdata}, we specify the model for the missing covariates as follows. Note that we remove from the notation the dependence on row number.

\begin{itemize}
\item $f(x_{\text{sex}} \mid p_{\text{sex}}) = \text{Bernoulli}(p_{\text{sex}})$,
\item $f(x_{\text{BMI}} \mid \xi_{\text{BMI}}, \omega_{\text{BMI}}, \alpha_{\text{BMI}}) = \text{Skew Normal}(\xi_{\text{BMI}}, \omega_{\text{BMI}}, \alpha_{\text{BMI}})$
\item $f(x_{\text{model year}} \mid \mu_{\text{model year}}, \sigma^2_{\text{model year}}) = \text{Normal}(\mu_{\text{model year}}, \sigma^2_{\text{model year}})$
\item $f(x_{\text{dvtotal}} \mid x_{\text{dvc}}, \alpha_{\text{dvtotal}}, \beta_{\text{dvtotal}}, \sigma^2_{\text{dvtotal}} ) = \text{Normal}(\alpha_{\text{dvtotal}}, + \beta_{\text{dvtotal}} x_{\text{DVC}}, \sigma^2_{\text{dvtotal}})$
\item $f(x_{\text{DVC}} \mid n_{\text{DVC}}, p_{\text{DVC}}) = \text{Neg-Binomial}(n_{\text{DVC}}, p_{\text{DVC}})$
\end{itemize}

So, connecting this new notation back with our old, the parameters used in the distributions of the covariates would be written out as $\bm{\alpha} = (p_{\text{sex}}, \xi_{\text{BMI}}, \omega_{\text{BMI}}, \alpha_{\text{BMI}}, \mu_{\text{model year}}, \sigma^2_{\text{model year}}, \alpha_{\text{dvtotal}}, \beta_{\text{dvtotal}}, \sigma^2_{\text{dvtotal}}, n_{\text{DVC}}, p_{\text{DVC}})$.

To specify the priors, we assume all elements of $\bm{\alpha}$ and $\bm{\beta}$ are independent a priori. This requires us only to specify marginal distributions for each parameter. The priors for the elements of $\bm{\alpha}$ are as follows:

\begin{itemize}
\item $f(p_{\text{sex}}) = \text{Beta}(50, 50)$
\item $f(\xi_{\text{BMI}}) = \text{Normal}(25, .5)$
\item $f(\omega_{\text{BMI}}) = \text{Log-Normal}(3.5, .1)$
\item $f(\alpha_{\text{BMI}}) = \text{Inverse-Gamma}(5,5)$
\item $f(\mu_{\text{model year}}) = \text{Normal}(-0.25, .001)$
\item $f(\sigma^2_{\text{model year}}) = \text{Inverse-Gamma}(200, 5970)$
\item $f(\alpha_{\text{dvtotal}}) = \text{Normal}(12,9)$
\item $f(\beta_{\text{dvtotal}}) = \text{Normal}(0,3)$
\item $f(\sigma^2_{\text{dvtotal}}) = \text{Inverse-Gamma}(1, .00005)$
\item $f(n_{\text{DVC}}) = \text{Inverse-Gamma}(94,114)$
\item $f(p_{\text{DVC}}) = \text{Inverse-Gamma}(.9, 59)$.
\end{itemize}

The priors for the elements of $\bm{\beta}$ are as follows:

\begin{itemize}
\item $f(\beta_{\text{int}}) = \text{Normal}(-5,3)$
\item $f(\beta_{\text{age}}) = \text{Normal}(.02, .02)$
\item $f(\beta_{\text{sex}}) = \text{Normal}(-.5, .4)$
\item $f(\beta_{\text{BMI}}) = \text{Normal}(0, .05)$
\item $f(\beta_{\text{dvtotal}}) = \text{Normal}(.006, .03)$
\item $f(\beta_{\text{kneebagdeploy}}) = \text{Normal}(0,.05)$
\item $f(\beta_{\text{SUV}}) = \text{Normal}(0,.3)$
\item $f(\beta_{\text{truck}}) = \text{Normal}(0,.3)$
\item $f(\beta_{\text{van}}) = \text{Normal}(0,.3)$
\item $f(\beta_{\text{modelyr}}) = \text{Normal}(0,.1)$
\item $f(\beta_{\text{farside}}) = \text{Normal}(0,.2)$
\item $f(\beta_{\text{nearside}}) = \text{Normal}(.2,.2)$.
\end{itemize}

Regarding the importance sampling proposal distribution, we choose 
$$
q_{\text{IS}}( \mathbf{x}_{\text{mis}} \mid \mathbf{x}_{\text{obs}}, \mathbf{y}) = q_{\text{IS}}(x_{\text{sex}})q_{\text{IS}}(x_{\text{BMI}}) q_{\text{IS}}(x_{\text{modelyr}}) q_{\text{IS}}(x_{\text{dvc}}) q_{\text{IS}}(x_{\text{dvt}} )
$$ 

where

\begin{itemize}
\item $q_{\text{IS}}(x_{\text{sex}}) = \text{Binomial}(1, .5)$
\item $q_{\text{IS}}(x_{\text{BMI}}) = t_{2}(20, 1)$ 
\item $q_{\text{IS}}(x_{\text{modelyr}}) = t_{20}(0,5)$ 
\item $q_{\text{IS}}(x_{\text{dvc}}) = \text{Neg-Binomial}(1.23, .015)$ 
\item $q_{\text{IS}}(x_{\text{dvt}} )) = \text{Log-Normal}(3, .5)$
\end{itemize}

Note we are using the Negative-Binomial parameterization that is used in base \verb|R|'s \verb|dnbinom| function: $\text{Neg-Binomial}(x ; n, p) = \frac{\Gamma(x+n)}{\Gamma(n) x!} p^n (1-p)^x$.

\begin{table}[ht]
\centering
\begin{tabular}{rrrrrr}
  \hline
 & Est. & MCSE & 95.cred.lower & 95.cred.upper & RHat \\ 
  \hline
  $p_{\text{sex}}$ & 0.925 & 0.001 & 0.901 & 0.957 & 1.468 \\ 
  $\xi_{\text{BMI}}$ & 24.172 & 0.015 & 23.589 & 24.697 & 1.001 \\ 
  $\omega_{\text{BMI}}$ & 6.770 & 0.130 & 5.532 & 11.652 & 1.443 \\ 
  $\alpha_{\text{BMI}}$ & 0.851 & 0.027 & 0.540 & 2.767 & 1.325 \\ 
  $\mu_{\text{model year}}$ & 12.734 & 0.027 & 11.233 & 13.287 & 1.001 \\ 
  $\sigma^2_{\text{model year}}$ & 0.048 & 0.002 & 0.006 & 0.084 & 1.007 \\ 
  $\alpha_{\text{dvtotal}}$ & 126.924 & 2.717 & 0.002 & 168.090 & 1.764 \\ 
  $\beta_{\text{dvtotal}}$ & -0.169 & 0.024 & -0.549 & 0.523 & 1.956 \\ 
  $\sigma^2_{\text{dvtotal}}$ & 198.008 & 4.291 & 72.836 & 304.664 & 1.344 \\ 
  $n_{\text{DVC}}$ & 0.394 & 0.011 & 0.149 & 0.554 & 1.070 \\ 
  $p_{\text{DVC}}$ & 0.002 & 0.000 & 0.001 & 0.009 & 1.364 \\ 
  $\phi$ & -2.819 & 0.007 & -3.129 & 0.352 & 1.035 \\ 
   \hline
\end{tabular}
\caption{Estimates for $\bm{\alpha}$ and $\phi$, along with Monte Carlo Standard Error Estimates, 95\% credible intervals, and Gelman-Rubin statistics.} \label{tab:alpha_phi_table}
\end{table}

\bibliographystyle{Chicago}
\bibliography{mybib}

\end{document}